\documentclass[aps,prb,preprint,showpacs,superscriptaddress,amsmath,amssymb]{revtex4}

\usepackage{graphicx}% Include figure files
\usepackage{dcolumn}% Align table columns on decimal point
\usepackage{bm}% bold math
\usepackage{color}% bold math

\usepackage{amssymb, amsmath}
\usepackage[T2A]{fontenc}
\usepackage[cp1251]{inputenc}

\begin{document}

\title{Self-induced THz-waves transmissivity of waveguides with finite-length layered superconductors}

\author{T.N.~Rokhmanova}
\affiliation{ A.Ya.~Usikov Institute for Radiophysics and Electronics, National Academy of Sciences of Ukraine, 61085 Kharkov, Ukraine}

\author{S.S.~Apostolov}
\affiliation{ A.Ya.~Usikov Institute for Radiophysics and
Electronics, National Academy of Sciences of Ukraine, 61085
Kharkov, Ukraine} \affiliation{V.N.~Karazin Kharkov National
University, 61077 Kharkov, Ukraine}

\author{Z.A.~Maizelis}
\affiliation{ A.Ya.~Usikov Institute for Radiophysics and
Electronics, National Academy of Sciences of Ukraine, 61085
Kharkov, Ukraine} \affiliation{V.N.~Karazin Kharkov National
University, 61077 Kharkov, Ukraine}

\author{V.A.~Yampol'skii}
\affiliation{ A.Ya.~Usikov Institute for Radiophysics and Electronics, National Academy of Sciences of Ukraine, 61085 Kharkov, Ukraine}
\affiliation{V.N.~Karazin Kharkov National University, 61077 Kharkov, Ukraine}
\affiliation{Advanced Science Institute, RIKEN, Saitama 351-0198, Japan}

\author{Franco Nori}
\affiliation{Advanced Science Institute, RIKEN, Saitama 351-0198, Japan}
\affiliation{Department of Physics, University of Michigan, Ann Arbor, Michigan 48109, USA}

\begin{abstract}
We predict and study theoretically  a new nonlinear electromagnetic phenomenon in a sample of layered superconductor of finite length placed in a waveguide with ideal walls. Two geometries are considered here: when the superconducting layers are parallel or perpendicular to the waveguide axis. We show that the transmittance of the superconductor slab can vary over a wide range, from nearly zero to one, when changing the amplitude of the incident wave. Thus, one can induce the total transmission or reflection of the incident wave by just changing its amplitude. Moreover, the dependence of the superconductor transmittance on the incident wave amplitude has a hysteretic behavior with jumps. The considered phenomenon of self-induced transparency can be observed even at small amplitudes, if the wave frequency $\omega$ is close to the cutoff frequency for linear Josephson plasma waves.
\end{abstract}


%74.78.Fk Multilayers, superlattices, heterostructures

% 74.50.+r -- Tunneling phenomena; Josephson effects

%74.72.-h Cuprate superconductors

\maketitle

\section{Introduction}

There is considerable interest in studies of metamaterials and nanostructures with unusual electromagnetic properties. Layered high-temperature superconductors, e.g., $\rm Bi_2Sr_2CaCu_2O_{8+\delta}$ single crystals, definitely belong to this type of materials. Experimental studies (see, e.g., Refs.~\onlinecite{Kl-Mu,Kl-Mu2}) have shown that the electrodynamics of layered superconductors can be described by a theoretical model, which assumes that thin superconducting $\rm CuO_2$ layers (with a thickness $s$ of about 2-3~{\AA}) are coupled through thicker dielectric layers (with a thickness $d$ of about 15~{\AA} and a dielectric constant $\varepsilon \sim 15$) via an \emph{intrinsic Josephson effect}. Due to the layered structure of $\rm Bi_2Sr_2CaCu_2O_{8+\delta}$ and similar compounds, they support the propagation of specific electromagnetic waves, the so-called Josephson plasma waves (JPWs) (see, e.g., Refs.~\onlinecite{Thz-rev,rev2} and references therein). These waves belong to the terahertz frequency range, which is very important for various applications but not easily accessible with modern electronic and optical devices. This technological perspective provides a strong motivation for research on these waves. From a scientific point of view, the interest in layered superconductors is mainly related to the specific type of plasma formed, the so-called Josephson plasma. An essential property of the Josephson plasma is the strong anisotropy of its current-carrying capability. This anisotropy manifests not only in the difference of the absolute values of the flowing currents (currents along the crystallographic \textbf{ab}-plane are two orders of magnitude higher than those along the \textbf{c}-axis) but also in their physical nature. Indeed, the current along the layers has the same nature as in the usual bulk superconductors and can be described within the framework of London's theory, while the current along the \textbf{c}-axis is of Josephson origin.

\subsection{Waves in usual plasmas versus Josephson plasma waves}

In the Josephson plasma, not only phenomena common to other types of plasmas can be observed but also those specific for layered superconductors. As in usual plasmas, there is a gap in the spectrum of Josephson waves. JPWs can only propagate with frequencies higher than the threshold Josephson plasma frequency $\omega_J$. As was theoretically demonstrated in Refs.~\onlinecite{surf,neg-ref}, the surface Josephson plasma waves (SJPWs) can propagate along the interface between layered superconductors and vacuum, similarly to usual plasmas. The excitation of these waves leads to various resonant phenomena~\cite{PhysRevB1,neg-ref}  similar to the Wood anomalies well known in optics (see, e.g., Refs.~\onlinecite{agr,wood,Petit}). However, contrary to usual plasmas, SJPWs can propagate with frequencies not only below the plasma frequency but also above it~\cite{neg-ref}. The Josephson plasma can also exhibit properties characteristic for left-handed media: a negative refractive index for terahertz waves can be observed at its interface with vacuum~\cite{neg-ref,negref}. As was shown in Ref.~\onlinecite{filter}, phenomena similar to Anderson localization and the formation of a transparency window for terahertz waves can be observed in layered superconductors with a randomly-fluctuating value of the maximum Josephson current.

\subsection{Nonlinearities}

Since the current along the \textbf{c}-axis is of Josephson origin, the electrodynamic equations for layered superconductors are nonlinear. This can lead to a number of non-trivial nonlinear effects accompanying the propagation of JPWs, e.g., slowing down of light~\cite{nl1}, self-focusing of terahertz pulses~\cite{nl1,nl2}, excitation of nonlinear waveguide modes~\cite{nl3}, as well as self-induced transparency of the slabs of layered superconductors and hysteretic jumps in the dependence of the slab transparency on the wave amplitude~\cite{nl4}.

\subsection{Finite-size samples}

It should be noted that the majority of the theoretical studies of JPWs consider infinite samples. However, the sizes of layered superconducting samples used in experiments are comparable or even smaller than the wavelength of the terahertz radiation. Obviously, in this case the sample cannot be treated as infinitely large. This means that the theory should consider finite sample sizes.

In this paper we discuss a nonlinear phenomenon that can be observed in a \emph{finite} slab of layered superconductor placed inside of a rectangular waveguide with ideal walls. We have considered two geometries: when waves are propagating across the layers or along them.  It turns out that the transmittance of the slab, being exposed by an incident wave from one of its sides, depends not only on the wave frequency, but also on the wave \emph{amplitude} for both of the geometries considered here. Therefore, a slab of fixed length can be completely transparent (when neglecting dissipation) for waves of certain amplitudes, and nearly totally reflecting for other amplitudes. Moreover, the dependence of the transmittance on the wave amplitude shows hysteretic behavior with jumps. The transmittance of the slab with layers perpendicular to the waveguide axis can vary over a wide range, from nearly zero to one, if the frequency $\omega$ of the irradiation is close to the Josephson plasma frequency $\omega_J$. Meanwhile, in the case of layers parallel to the waveguide axis, the transmittance can change significantly at frequencies far from $\omega_J$.

The paper is organized as follows. In the next section, we discuss the configuration where waves propagate in a waveguide along the superconducting layers. We start with the geometry of the problem and present the main equations for the electromagnetic fields both in the vacuum regions of the waveguide and also in the slab of the layered superconductor. The electrodynamics of layered superconductors is described by nonlinear coupled sine-Gordon equations~\cite{sine-gord,SG2,SG3,SG4,SG5,Thz-rev}. The nonlinearity originates from the nonlinear relation $J\propto\sin\varphi$ between the Josephson interlayer current $J$ and the gauge-invariant interlayer phase difference $\varphi$ of the order parameter. We emphasize that the nonlinearity can play a crucial role in the propagation of JPWs even for small wave amplitudes, $|\varphi| \ll 1$, when $\sin\varphi$ can be expanded into a series: $\sin\varphi \approx \varphi-\varphi^3/6$. Then we express the transmittance $T$ in terms of the amplitude of the incident wave and analyze this dependence for different frequency ranges. In the third section, we discuss the same points for the second configuration, when waves propagate in a waveguide across the superconducting layers.

\section{Propagation of waves along the superconducting layers}\label{ab}

\subsection{Geometry of the problem}

\begin{figure}[h]
\begin{center}
\includegraphics [width=14cm]{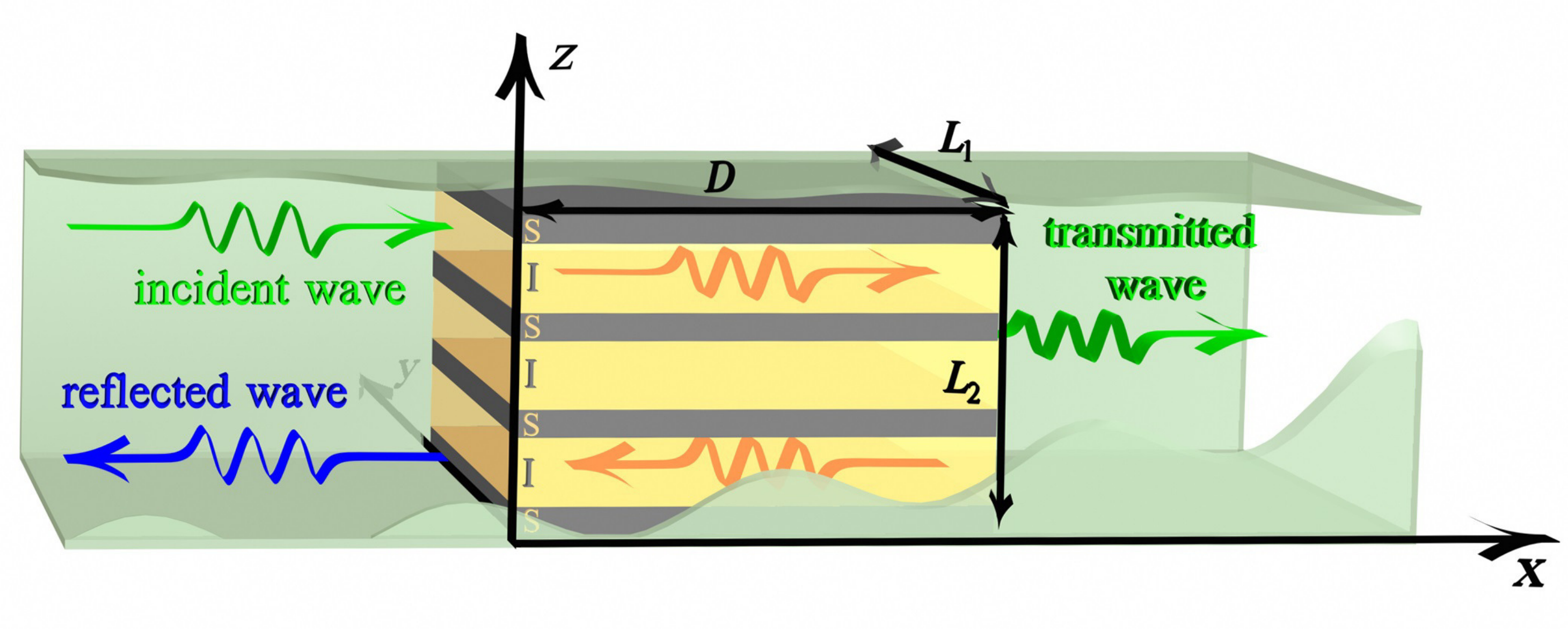}
\caption{(Color online) Schematic geometry for JPWs propagating in a  waveguide along the superconducting layers. Note that here S and I stand for superconducting and insulator layers, respectively. The green translucent layer (cut-off to show the sample inside) represents the walls of the waveguide.} \label{wavegAB}
\end{center}
\end{figure}

Consider a waveguide of lateral sizes $L_1$ and $L_2$ with a slab of layered superconductor of length $D$ inside it. The coordinate system is chosen in such a way that the crystallographic $\mathbf{ab}$-plane of the layered superconductor coincides with the $xy$-plane, and the $\mathbf{c}$-axis is along the $z$-axis. We start our consideration from the case when the electromagnetic wave of frequency $\omega$ propagates in the waveguide along the $x$-axis, which is parallel to the superconducting layers (see Fig.~\ref{wavegAB}).

We now consider an extraordinary incident wave with the magnetic field parallel to the superconducting layers,
\begin{equation}\label{polarization}
{\vec E} = \{0, 0, E_z\}, \qquad {\vec H} = \{H_x, H_y, 0\}.
\end{equation}
For simplicity, we assume here that the electromagnetic field is uniform in the $z$-direction (across the layers) and the electric field is perpendicular to the superconducting layers. Otherwise, the transformation of the polarization of the wave would occur after its reflection and transmission through the anisotropic superconductor. The incident wave is partly reflected and partly transmitted through the slab.

\subsection{Electromagnetic field in the vacuum regions}\label{vacuum_ab}

The waveguide has two vacuum regions (see Fig.~\ref{wavegAB}). In the first one (at $x<0$), the electromagnetic field can be represented as a sum of incident and reflected waves with amplitudes $H_i$ and $H_r$, respectively. Using the Maxwell equations and boundary conditions (the equality of the tangential components of the electric field to zero on the waveguide walls), one can derive the electric and magnetic field components,
\begin{gather}
E_z^{(v1)}=\left[H_i \cos(k_x x-\omega t)\right.
\notag\\
\left. - H_r \cos(k_xx+\omega t+\alpha)\right]\sin(k_yy),
\notag\\
H_x^{(v1)}=-\frac{k_y}{k}\left[H_i \sin(k_xx-\omega t) \right.
\notag\\
\left. + H_r \sin(k_xx+\omega t+\alpha)\right]\cos(k_yy),
\label{vacuum1ab}\\
H_y^{(v1)}=\frac{k_x}{k}\left[H_i \cos(k_xx-\omega t) \right.
\notag\\
\left. + H_r \cos(k_xx+\omega t+\alpha)\right]\sin(k_yy).
\notag
\end{gather}
Here
\begin{equation}
k_y=\frac{n\pi}{L_1}, \qquad k_x= \big(k^2-k_y^2\big)^{1/2}, \qquad k=\frac{\omega}{c},
\end{equation}
$n$ is a positive integer number that defines the propagating modes in the waveguide, $\alpha$ is the phase shift of the reflected wave, and $c$ is the speed of light.

There is only a transmitted wave of amplitude $H_t$ in the second vacuum region (at $x> D$). Here the electromagnetic field components  are
\begin{gather}
E_z^{(v2)}= H_t \cos[k_x(x-D)-\omega t+\beta]\sin(k_yy),
\notag\\
H_x^{(v2)}=-\frac{k_y}{k} H_t \sin[k_x(x-D)-\omega t+\beta]\cos(k_yy),
\label{vacuum2ab}\\
H_y^{(v2)}=\frac{k_x}{k} H_t \cos[k_x(x-D)-\omega t+\beta]\sin(k_yy)
\notag
\end{gather}
where $\beta$ is the phase shift of the transmitted wave.

\subsection{Electromagnetic field in the slab of layered superconductor}\label{sconductor_ab}

The electromagnetic field in a slab of layered superconductor  is related to the distribution of the gauge-invariant interlayer phase difference $\varphi(x,y,z,t)$ of the order parameter (see, e.g., Ref.~\onlinecite{Thz-rev}). For the extraordinary wave, this  relation is defined by the equations,
\begin{gather}
E_x^s=-\frac{\lambda_{ab}^2}{c} \frac{\partial^2 H_y}{\partial t \partial z},\, E_y^s=\frac{\lambda_{ab}^2}{c} \frac{\partial^2 H_x}{\partial t \partial z},\,
E_z^s=\mathcal{H}_0 \frac{1}{\omega_J \sqrt{\varepsilon}} \frac{\partial \varphi}{\partial t},
\notag\\
\label{s-fields}
\frac{ \partial H_y^s}{\partial x} - \frac{\partial H_x^s}{\partial y}= \frac{\mathcal{H}_0}{\lambda_c} \left[\sin\varphi + \frac{1}{\omega_J^2}\frac{\partial^2 \varphi}{\partial t^2} \right].
\end{gather}
Here ${\cal H}_0=\Phi_0/2\pi d\lambda_c$, $\Phi_0=\pi c\hbar/e$ is the magnetic flux quantum, $\lambda_{ab}$ and $\lambda_c=c/\omega_J\varepsilon^{1/2}$ are the London penetration depths across and along the layers, respectively, $\omega_J = (8\pi e d J_c/\hbar\varepsilon)^{1/2}$ is the Josephson plasma frequency, $J_c$ is the maximal value of the Josephson current density, and $e$ is the elementary charge.  We do not take into account the relaxation terms because they are small at low temperatures and do not play an essential role in the phenomena considered here.

The phase difference is defined by a solution of a set of coupled sine-Gordon equations that, in the continuous limit, has the following form (see, e.g., Ref.~\onlinecite{Thz-rev} and references therein):
\begin{equation}\label{sine-Gordon}
\left(1-\lambda_{ab}^2\frac{\partial^2}{\partial
z^2}\right)\left[\frac{1}{\omega_J^2}\frac{\partial^2
\varphi}{\partial t^2} + \sin\varphi\right]-
\lambda_c^2\left(\frac{\partial^2 \varphi}{\partial x^2}+\frac{\partial^2 \varphi}{\partial y^2}\right)=0.
\end{equation}
Note that the $z$-component of the electric field causes the breakdown of electro-neutrality of superconducting layers. This results in an additional, so-called capacitive, interlayer coupling.  This coupling can significantly affect  the properties of the \emph{longitudinal} JPWs with wave vectors oriented across the layers. The dispersion equation for linear plane JPWs with account of the capacitive coupling was obtained in Ref.~\onlinecite{helm3}. According to this dispersion equation, the capacitive coupling is important if the component $k_z$ (normal to the layers) is close to $k=\omega/c$. For the cases when $(k_x^2+k_y^2)^{1/2}= \omega/c$, $k_z=0$ (considered in this section) or $(k_x^2+k_y^2)^{1/2}\sim k_z \sim \omega/c$ (considered in the next section),  the capacitive coupling can be safely neglected due to the smallness of the parameter $\alpha = R_D^2\varepsilon/sd$. Here $R_D$ is the Debye length for a charge in a superconductor.

For the mode in Eq.~\eqref{polarization} with an electric field $E_z^s$ proportional to $\sin(k_yy)$ and uniform in the $z$-direction, Eq.~\eqref{sine-Gordon} is rewritten as
\begin{equation}\label{sine-Gordon-new}
\left(\frac{1}{\omega_J^2}\frac{\partial^2
\varphi}{\partial t^2} + k_y^2\lambda_c^2 \varphi + \sin\varphi\right) -
\lambda_c^2\frac{\partial^2 \varphi}{\partial x^2}=0.
\end{equation}
Equation \eqref{sine-Gordon-new} shows that a \emph{linear} Josephson plasma wave can propagate along the waveguide if its frequency exceeds the cutoff frequency $\omega_{\rm cut}$,
\begin{equation}\label{omega-cut}
\omega_{\rm cut} = \omega_J (1+k_y^2 \lambda_c^2)^{1/2}.
\end{equation}
Here we consider the weakly nonlinear waves when the Josephson current density $J_c \sin \varphi$ can be replaced approximately by  $ J_c(\varphi-\varphi^3/6)$. We assume that the wave frequency $\omega$ is close to $\omega_{\rm cut}$. In this case, the linear terms in brackets in Eq.~\eqref{sine-Gordon-new} nearly cancel each other. Therefore, in spite of the weakness of the nonlinearity, the term $\varphi^3$ plays a very important role in the wave propagation. Moreover, one can neglect the generation of higher harmonics when the frequency $\omega$ is close to $\omega_{\rm cut}$ (similarly to the effects considered in Refs.~\onlinecite{nl1,nl3}).

We seek a solution of Eq.~\eqref{sine-Gordon-new} in the form of a wave with the $x$-dependent amplitude $a$ and phase $\eta$,
\begin{equation}\label{varphi_ab}
\varphi (x, y, t) = a(x)\mu\kappa \sin[\eta(x)-\omega t]\sin(k_yy),
\end{equation}
where
\begin{equation}\label{kappa_pm}
\mu=\frac{8}{3}\sqrt{2},\, \kappa=|\Omega^2-\Omega_{\rm cut}^2|^{1/2}, \, \Omega=\frac{\omega}{\omega_J}, \, \Omega_{\rm cut}=\frac{\omega_{\rm cut}}{\omega_J}.
\end{equation}
Introducing the dimensionless coordinate and the normalized length of the sample,
\begin{equation}\label{notations_ab}
\xi=\frac{x}{\lambda_c}\kappa\,, \qquad \delta=\frac{D}{\lambda_c}\kappa\,,
\end{equation}
and substituting the phase difference $\varphi$ given by Eq.~\eqref{varphi_ab} into Eq.~\eqref{sine-Gordon-new}, we obtain two differential equations for the functions $\eta(\xi)$ and $a(\xi)$,
\begin{equation}\label{from_sine-Gordon_ab-1}
\eta^\prime(\xi)=-\frac{L}{a^2(\xi)},
\end{equation}
\begin{equation}\label{from_sine-Gordon_ab-2}
a^{\prime\prime}(\xi)=- \sigma a(\xi)-a^3(\xi)+\frac{L^2}{a^3(\xi)}.
\end{equation}
Here $\sigma = {\rm sign} (\Omega-\Omega_{\rm cut})$, $L$ is an integration constant,  and the prime denotes derivation over $\xi$.

Using Eqs.~(\ref{s-fields}) and \eqref{varphi_ab}, we express the tangential components of the electric and magnetic fields via the functions $a(\xi)$ and $\eta(\xi)$,
\begin{align}
&E_z=-\mathcal{H}_0\lambda_c\mu\kappa ka(\xi)\cos[\eta(\xi)-\omega t]\sin(k_yy),
\notag\\ \label{tan_components_ab}
&H_y=\mathcal{H}_0\mu\kappa^2 \left[a(\xi)\sin[\eta(\xi)-\omega t]\right]^\prime\sin(k_yy).
\end{align}
Now we can find the amplitudes of the reflected and transmitted waves by solving Eqs.~\eqref{from_sine-Gordon_ab-1}, \eqref{from_sine-Gordon_ab-2} and matching the tangential components of the electric and magnetic fields at both interfaces (at $x=0$ and $x=D$) between the vacuum regions and the layered-superconductor.

\subsection{Transmittance of the superconducting slab}

In this subsection, we analyze the transmittance $T$ of the finite-length slab of layered superconductor  in the waveguide. Matching the fields given by Eqs.~\eqref{tan_components_ab} for the superconductor and Eqs.~\eqref{vacuum1ab}, \eqref{vacuum2ab} for the vacuum regions, we  obtain the following three equations for the amplitudes $a(0)$, $a(\delta)$ and their derivatives on both edges of the layered superconductor, as well as the expression for the amplitude of the transmitted wave:
\begin{gather}
\left[\Gamma a^\prime(0)\right]^2+\left[\dfrac{\Gamma L}{a(0)}+a(0)\right]^2=4h_i^2,
\label{1ab}\\
a^2(\delta)=\Gamma L,
\label{2ab}\\
a^\prime(\delta)=0,
\label{3ab}\\
h_t^2=a^2(\delta).
\label{4ab}
\end{gather}
Here
\begin{equation}\label{hi_ab}
h_m=\dfrac{H_m}{\mathcal{H}_0k\lambda_c\mu\kappa},\quad m=i,t
\end{equation}
are the normalized amplitudes of the incident ($m=i$) and transmitted ($m=t$) waves, and $\Gamma=\kappa/(k_x\lambda_c)$.

Equations~(\ref{1ab}--\ref{3ab}) together with Eqs.~\eqref{from_sine-Gordon_ab-1} and \eqref{from_sine-Gordon_ab-2} determine the integration constant $L$ as a function of the amplitude $h_i$ of the incident wave. Using Eqs.~\eqref{2ab} and \eqref{4ab}, we find that the constant $L$ defines directly the transmittance $T$ of the superconducting slab:
\begin{equation}\label{R_ab}
T=\frac{h_t^2}{h_i^2} =\frac{a^2(\delta)}{h_i^2}=\frac{\Gamma }{h_i^2}L.
\end{equation}

As we show below, the dependence of the transmittance $T$ on the amplitude of the incident wave is \emph{multivalued} due to the nonlinearity of Eqs.~\eqref{from_sine-Gordon_ab-1} and \eqref{from_sine-Gordon_ab-2}. Below, we analyze this dependence for the cases of negative  and positive frequency detunings $(\Omega - \Omega_{\rm cut})$  (negative  and positive $\sigma$).

\subsubsection{Transmittance of a superconducting slab for negative frequency detunings ($\sigma = -1$) \label{sigma = -1}}

We start our consideration from the case of \emph{negative} frequency detunings when
\begin{equation}\label{small-omega}
\Omega < \Omega_{\rm cut}.
\end{equation}
As was mentioned above, the \emph{linear Josephson plasma waves cannot propagate} along the superconducting layers under such conditions, i.e.,  the transmittance of the slab of layered superconductor  is exponentially small because of the skin effect. However, the nonlinearity promotes the wave propagation due to the effective decrease of the cutoff frequency.

\begin{figure}
\begin{center}
\includegraphics*[width=14cm]{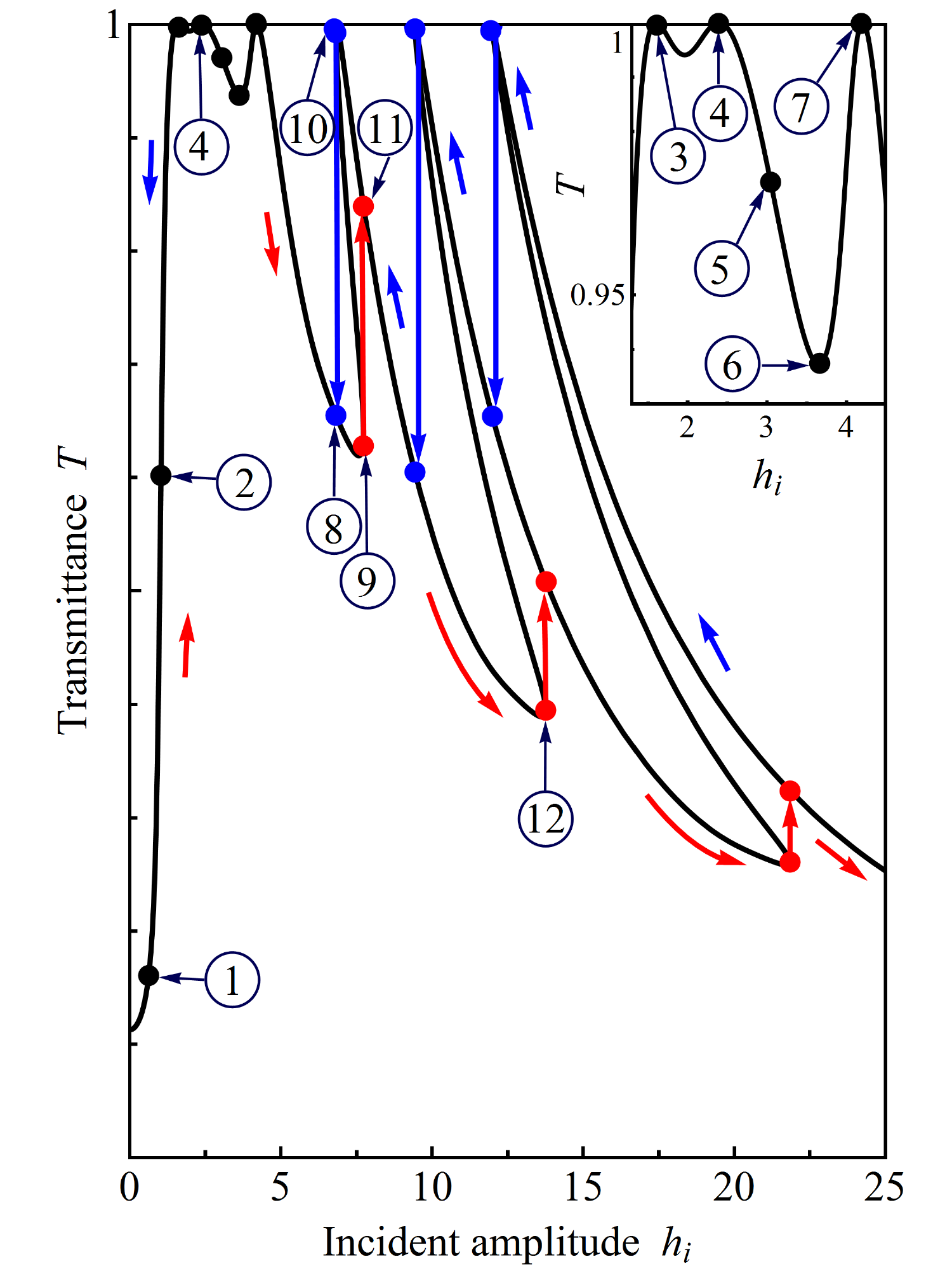}
\caption{(Color online) The transmittance $T$ versus the normalized amplitude $h_i$ of the incident wave for \emph{negative} frequency detuning, $\Omega - \Omega_{\rm cut}<0$.  Movement along the lower red (upper blue) arrows shows the evolution of the transmittance when increasing (decreasing) the amplitude $h_i$. The numbers near the points on the $T(h_i)$ curve correspond to the same numbers of the phase trajectories $a'(a)$ shown in Figs.~\ref{phase_diagr_ab-1} and~\ref{phase_diagr_ab-2}. The inset demonstrates the enlarged region near the first three maximums in the $T(h_i)$ dependence. The values of the parameters are: $\kappa= 0.1$, $\delta=1.5$ ($D=15\lambda_c$), $\lambda_c=4\cdot 10^{-3}$~cm, $\lambda_{ab}=2000$~\AA, $\omega_J/2\pi=0.3$~THz, $n_1=1$, and $L_1=0.1$~cm.} \label{T(hi)_ab-1}
\end{center}
\end{figure}

\begin{figure}
\begin{center}
\includegraphics*[width=14cm]{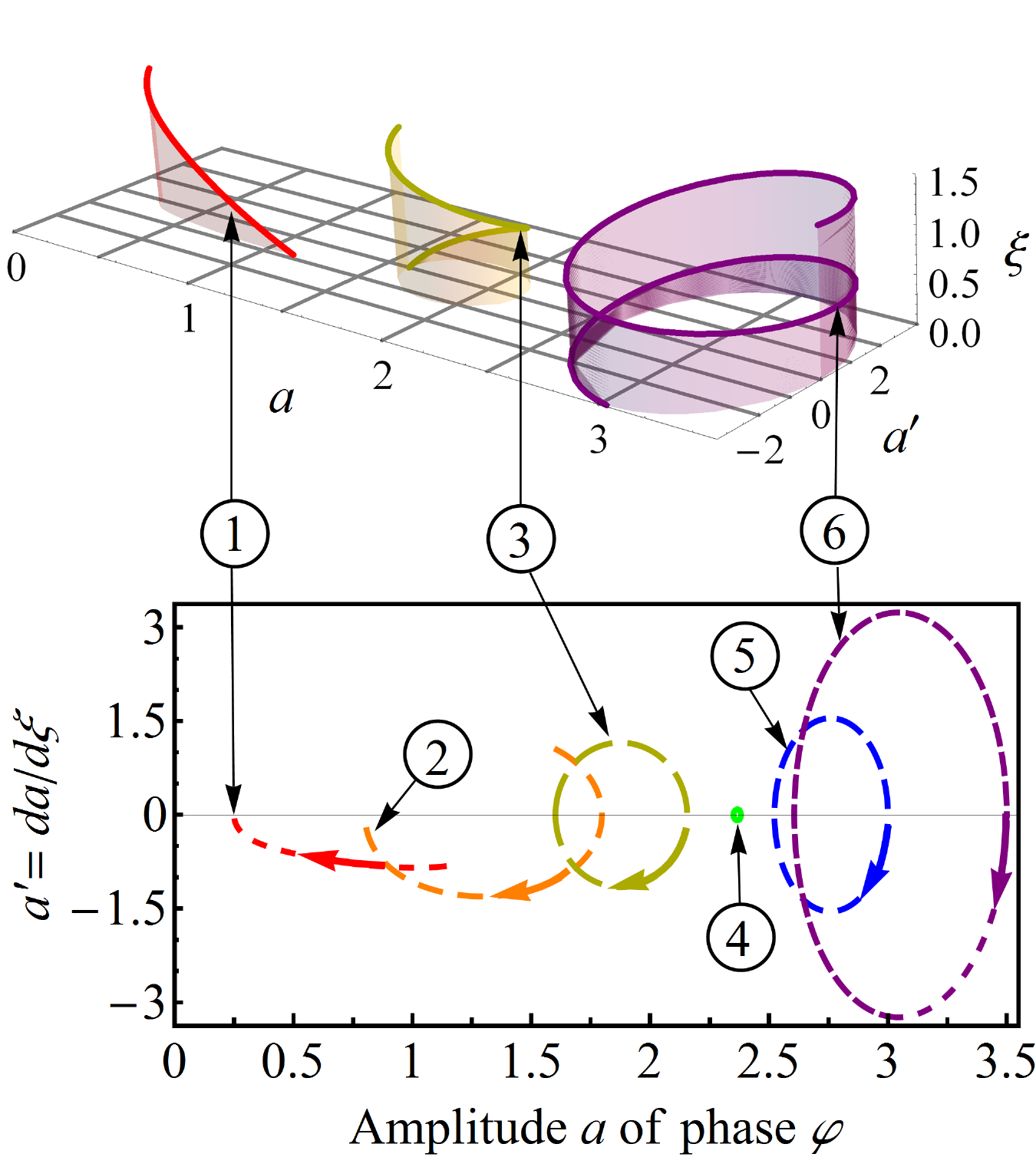}
\caption{(Color online) The phase trajectories $a'(a)$ for \emph{negative} frequency detuning, $\Omega - \Omega_{\rm cut}<0$. These trajectories correspond to points from 1 to 6 in the $T(h_i)$ plot shown in Fig.~\ref{T(hi)_ab-1}. The upper panel demonstrates in 3D the dependences of $a$ and $a'$ on the spatial coordinate $\xi$ along the slab (for points 1, 3, and 6 in Fig.~\ref{T(hi)_ab-1}). The parameters here are the same as in Fig.~\ref{T(hi)_ab-1}.}
\label{phase_diagr_ab-1}
\end{center}
\end{figure}

\begin{figure}
\begin{center}
\includegraphics*[width=14cm]{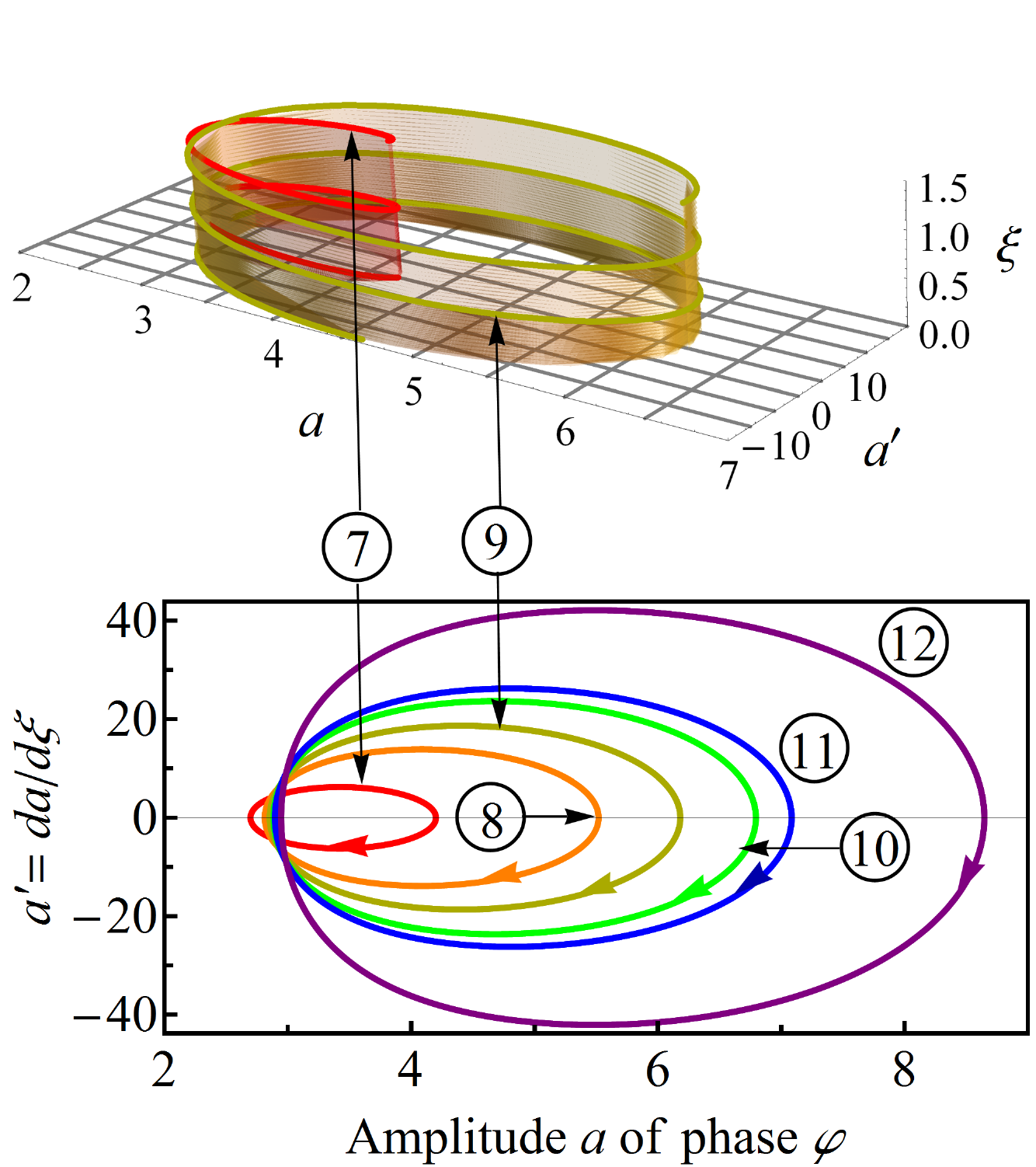}
\caption{(Color online) The same as in Fig.~\ref{phase_diagr_ab-1} for the points 7--12 shown also in Fig.~\ref{T(hi)_ab-1}.}
\label{phase_diagr_ab-2}
\end{center}
\end{figure}

Using Eqs.~\eqref{from_sine-Gordon_ab-1}, \eqref{from_sine-Gordon_ab-2}  and conditions (\ref{1ab}--\ref{3ab}), we find the integration constant $L$ and  the transmittance $T$ given by Eq.~\eqref{R_ab}. Figure~\ref{T(hi)_ab-1} shows the numerically-calculated dependence $T(h_i)$. We consider also the spatial distribution of the amplitude $a(\xi)$ and the phase trajectories $a'(a)$. The curves $a'(a)$ are shown in the bottom panels of Figs.~\ref{phase_diagr_ab-1} and \ref{phase_diagr_ab-2}. The movement along the spatial coordinate $\xi$ (proportional to $x$, as defined in Eq.~\eqref{notations_ab}) from zero to $\delta$ is shown by arrows. The upper panels in Figs.~\ref{phase_diagr_ab-1} and \ref{phase_diagr_ab-2} demonstrate the 3D curves for $a$ and $a'$ dependences on the coordinate $\xi$. Each trajectory corresponds to some value of the amplitude $h_i$. According to Eqs.~\eqref{2ab} and~\eqref{R_ab}, the value $a(\delta)$ defines the transmittance of the slab.

Consider in detail the dependence of the transmittance $T$ on the amplitude $h_i$ of the incident wave shown in Fig.~\ref{T(hi)_ab-1}. When \emph{increasing} the amplitude $h_i$, the transmittance grows following the red lower arrows and passing the points 1 and 2 in the $T(h_i)$ plot. For such amplitudes, the phase trajectories $a'(a)$ represent non-closed curves (see curves 1 and 2 in Fig.~\ref{phase_diagr_ab-1}). At $h_i \approx 1.7$ the phase trajectory turns into a  closed loop (curve 3 in Fig.~\ref{phase_diagr_ab-1}), and transmittance achieves its maximum  value $T = 1$. Then, when increasing $h_i$, the transmittance oscillates, achieving the maximum value $T=1$ in points 4 and 7 (in Fig.~\ref{T(hi)_ab-1}). The interesting thing is that there exists a special value of $h_i \approx 2.4$ where the phase trajectory $a'(a)$ shrinks into a point (see point 4 in Fig.~\ref{phase_diagr_ab-1}). This means that, for such value of $h_i$, the amplitude $a$ of the electromagnetic wave, given by Eq.~\eqref{tan_components_ab}, is constant inside the slab and the phase $\eta$ changes linearly with $\xi$, similarly to linear waves. The slab is completely transparent in this case (see point 4 in Fig.~\ref{T(hi)_ab-1}).

The continuous change of the transmittance with an increase of $h_i$ is abruptly terminated in the point 9 (at $h_i\approx7.7$) in Fig.~\ref{T(hi)_ab-1}. Further increase of the amplitude $h_i$ results in a jump to the upper branch of the $T(h_i)$ dependence (to the point 11). Similar jumps occur again and again when further increasing the amplitude $h_i$ of the incident wave.

Let us now consider the $T(h_i)$  dependence when \emph{decreasing} the incident wave amplitude $h_i$, following the blue upper arrows in Fig.~\ref{T(hi)_ab-1}. The decrease of $h_i$ causes an increase of the transmittance $T$ untill we achieve the situation when the slab is completely transparent. Here a branch of the $T(h_i)$  dependence breaks and further movement along the curve is possible only after a jump to the lower branch. As one can see from Fig.~\ref{T(hi)_ab-1}, an additional decrease of the amplitude $h_i$ results in a similar behavior until we reach the point 10 ($h_i\approx6.8$), where the last jump to the lower branch (to the point 8) occurs. Further decrease does not exhibit additional jumps.

Thus, for frequencies $\Omega $ less than the cutoff frequency $\Omega_{\rm cut}$, the transmittance of a superconducting slab has a hysteretic dependence with jumps on the amplitude of the incident wave.

\begin{figure}
\begin{center}
\includegraphics*[width=14cm]{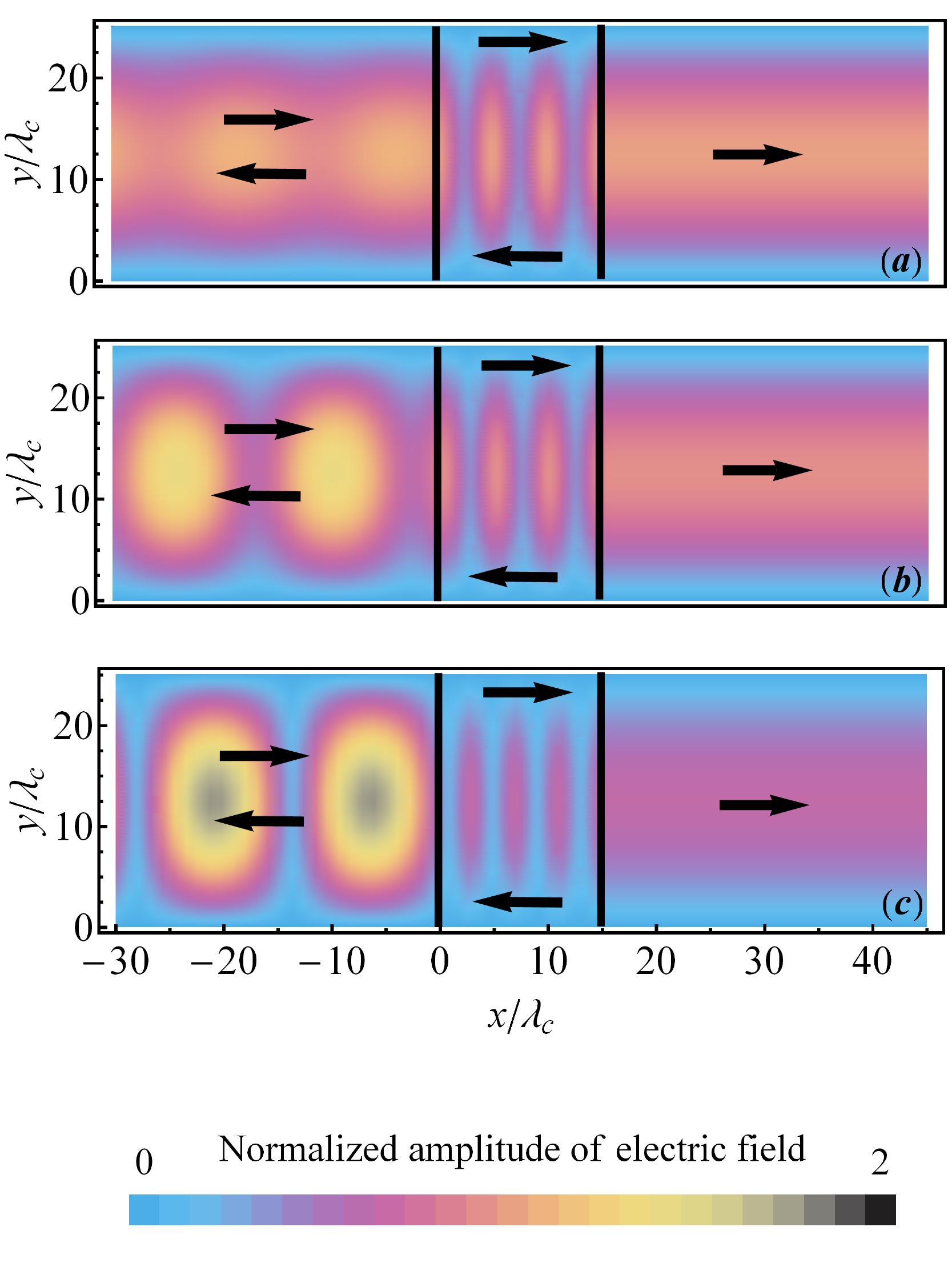}
\caption{(Color online) The spatial distribution of the electric field amplitude (normalized to the incident wave amplitude $H_i$) inside the waveguide. Panels (a), (b), and (c) correspond to the points 10, 11, and 12 in Fig.~\ref{T(hi)_ab-1}, respectively. The color determines the value of the amplitude. The vertical straight lines show the edges of the superconducting sample. The parameters are the same as in Fig.~\ref{T(hi)_ab-1}.}
\label{colorplot_ab}
\end{center}
\end{figure}

In Fig.~\ref{colorplot_ab}, we illustrate the spatial distribution of the electric field amplitude inside the waveguide for three values of the incident amplitude $h_i$. Panel \ref{colorplot_ab}(a) corresponds to the point 10 in Fig.~\ref{T(hi)_ab-1}, where the layered superconductor is almost transparent. This means that the reflected wave is almost absent and the interference pattern in the vacuum regions is too blurry. The panels \ref{colorplot_ab}(b) and \ref{colorplot_ab}(c) correspond to the points 11 and 12 in Fig.~\ref{T(hi)_ab-1}. They show a pronounced interference pattern for the region $x<0$. Note that the field distribution along the $y$-axis corresponds to $n_1=1$.

\subsubsection{Transmittance of a superconducting slab for  positive frequency detunings ($\sigma = +1$)\label{sigma = +1}}

\begin{figure}
\begin{center}
\includegraphics*[width=14cm]{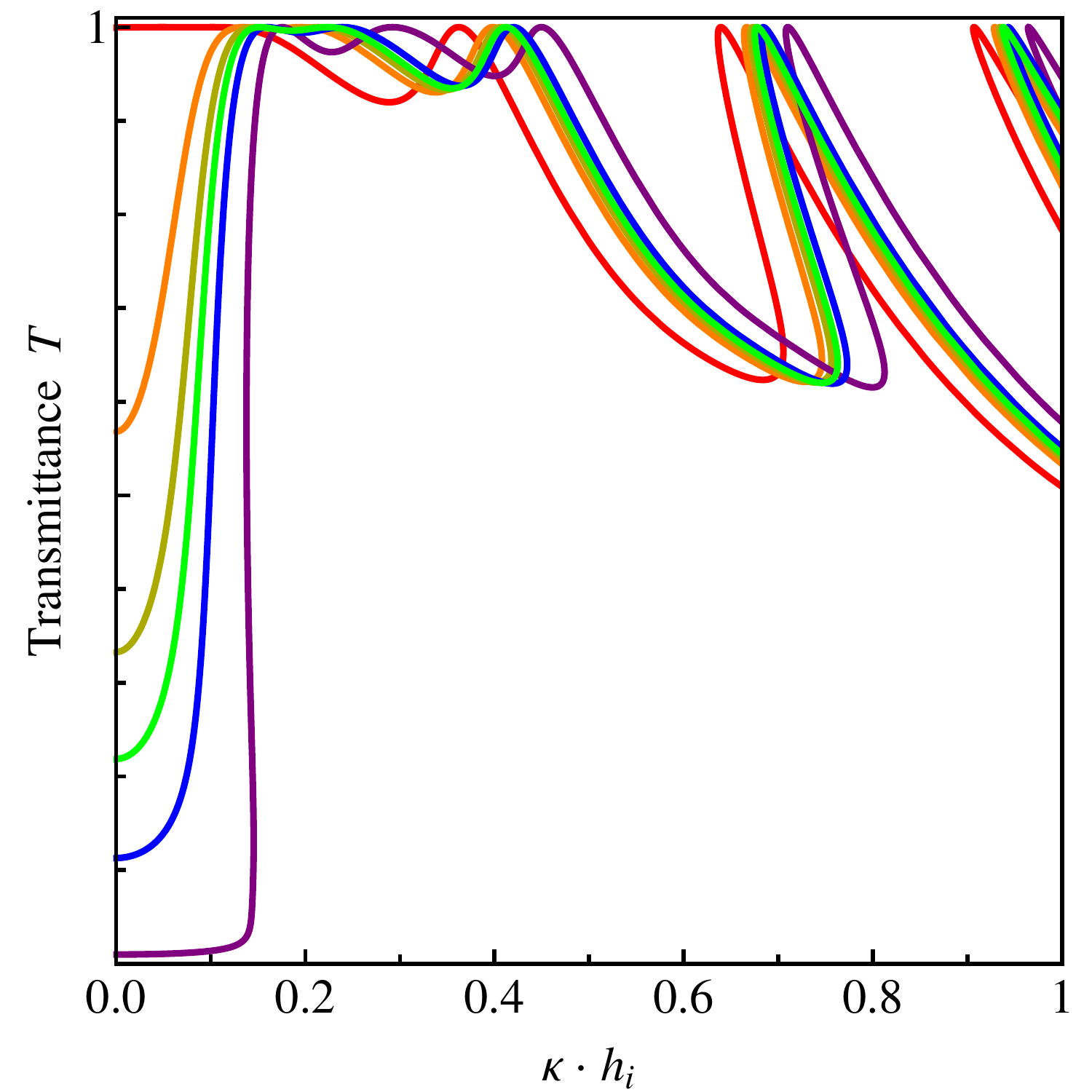}
\caption{(Color online) Dependence of the transmittance $T$ on the value of $h_i \cdot \kappa$ for different frequency detunings. According to Eq.~\eqref{hi_ab}, the argument $h_i \cdot \kappa$ is proportional to the non-normalized amplitude $H_i$ of the incident wave and is independent of the frequency detuning. The red, orange, and yellow curves (the upper curves at small amplitudes) correspond to positive frequency detunings with $\kappa= 0.2$, 0.1, and 0.05, respectively.  The purple, blue, and green curves (the lower curves at small amplitudes) correspond to negative frequency detunings with $\kappa= 0.2$, 0.1, and 0.05, respectively. Other parameters are the same as in Fig.~\ref{T(hi)_ab-1}.}
\label{fr6}
\end{center}
\end{figure}

Now we discuss the case of \emph{positive} frequency detunings when
\begin{equation}\label{large-omega}
\Omega > \Omega_{\rm cut}.
\end{equation}
Under this condition, contrary to the case $\Omega < \Omega_{\rm cut}$, even the linear Josephon plasma waves can propagate in the superconductor along the layers. Therefore, the transmittance is not exponentially small in the linear regime. It vary over a wide range, from zero to one, depending on the relation between the length of the slab and the wavelength. Figure~\ref{fr6} shows the dependence of the transmittance on the incident wave amplitude for different frequency detunings. For proper comparison of the curves, we use here another normalization of the incident wave amplitude, independent of $\kappa$ and, therefore, different from  Eq.~\eqref{hi_ab}. Namely, we plot all the curves as functions of the parameter $h_i \cdot \kappa$. One can see that the transmittance $T$ depends strongly on $\kappa$ and $\sigma$ for small values of $h_i \cdot \kappa$. However, all the dependences of $T$ on the incident wave amplitude show similar behavior at larger $h_i \cdot \kappa$. This means that, for high enough amplitudes $H_i$, the nonlinearity plays the most important role in the transmissivity,  regardless of the frequency detuning.

\subsection{Mechanical analogy}

The problem discussed here has an interesting mechanical analogy. Indeed, Eqs.~\eqref{from_sine-Gordon_ab-1} and \eqref{from_sine-Gordon_ab-2} coincide  with equations describing a motion of a particle of mass one in a centrally symmetric field. The amplitude $a(\xi)$, the phase $\eta(\xi)$, and the coordinate $\xi$ along the layers of the superconductor play the roles of the radial coordinate, polar angle, and time, respectively. The constant $L$ in Eqs.~\eqref{from_sine-Gordon_ab-1} and \eqref{from_sine-Gordon_ab-2} can be considered as the conserved angular momentum of the particle.

Integrating Eq.~\eqref{from_sine-Gordon_ab-2}, we find the energy conservation law,
\begin{equation}\label{energy_ab}
\dfrac{(a')^2}{2}+ U_{\rm eff}(a)={\cal E},
\end{equation}
with the effective potential energy
\begin{equation}\label{ueff_ab}
U_{\rm eff} (a)=\dfrac{L^2}{2a^2}+\sigma\dfrac{a^2}{2}+\dfrac{a^4}{4}.
\end{equation}
The first term in Eq.~\eqref{energy_ab} is the kinetic energy of the radial motion, ${\cal E}$ is the total energy, the first term in Eq.~\eqref{ueff_ab} is the centrifugal energy, and the last two terms in Eq.~\eqref{ueff_ab} describe the potential of the central field.
\begin{figure}
\begin{center}
\includegraphics*[width=14cm]{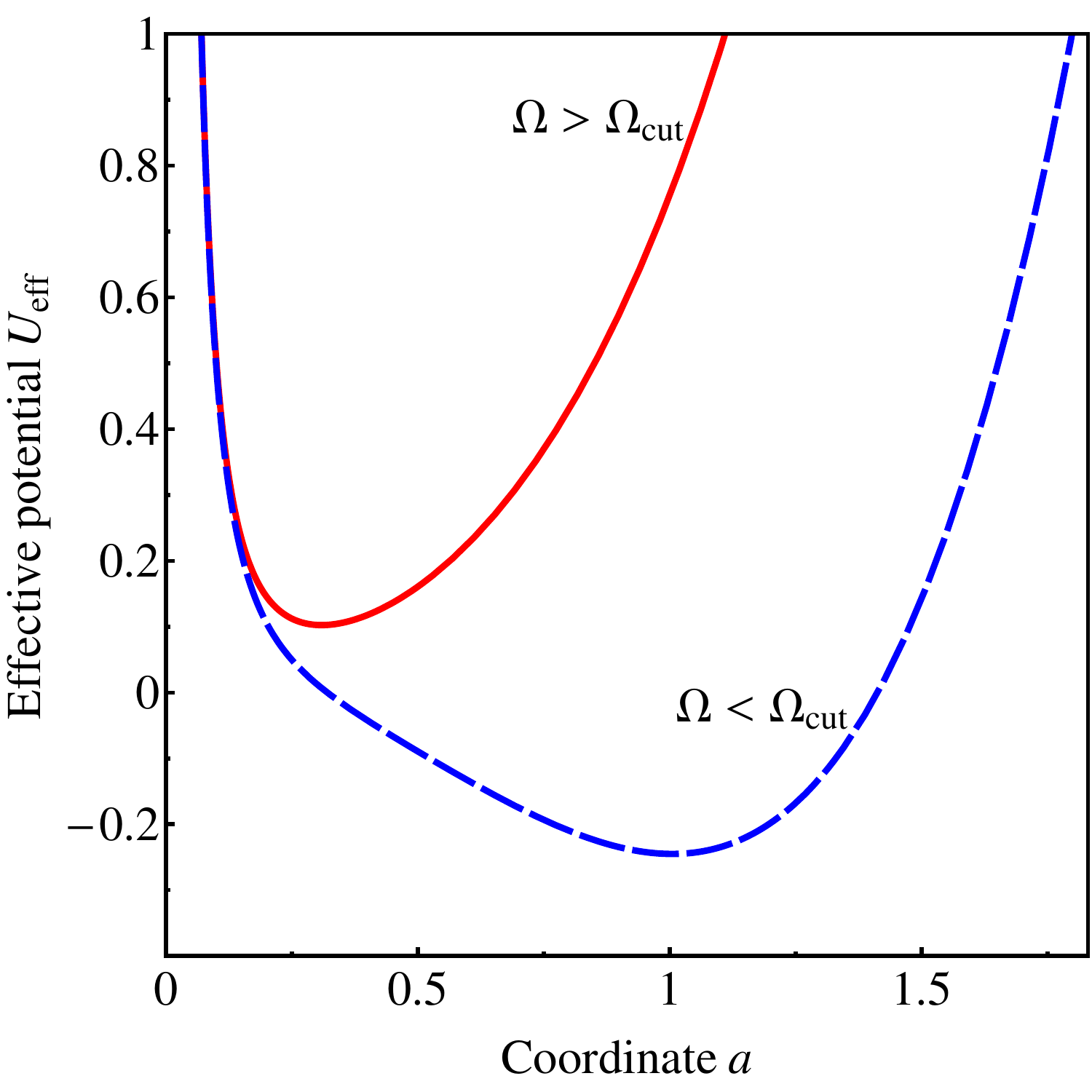}
\caption{(Color online) Dependence of the potential $U_{\rm eff}$ given by Eq.~\eqref{ueff_ab} on the radial coordinate $a$. The movement of the particle in this potential is the mechanical analog for the spatial distribution of the amplitude $a$ of the electric field in the superconductor. The solid red curve corresponds to the positive frequency detuning and the  dashed blue curve corresponds to the negative one. The value of the constant $L$ is $0.1$, and the other parameters are the same as in Fig.~\ref{T(hi)_ab-1}.} \label{mechanical_analogy_ab}
\end{center}
\end{figure}

\begin{figure}
\begin{center}
\includegraphics*[width=14cm]{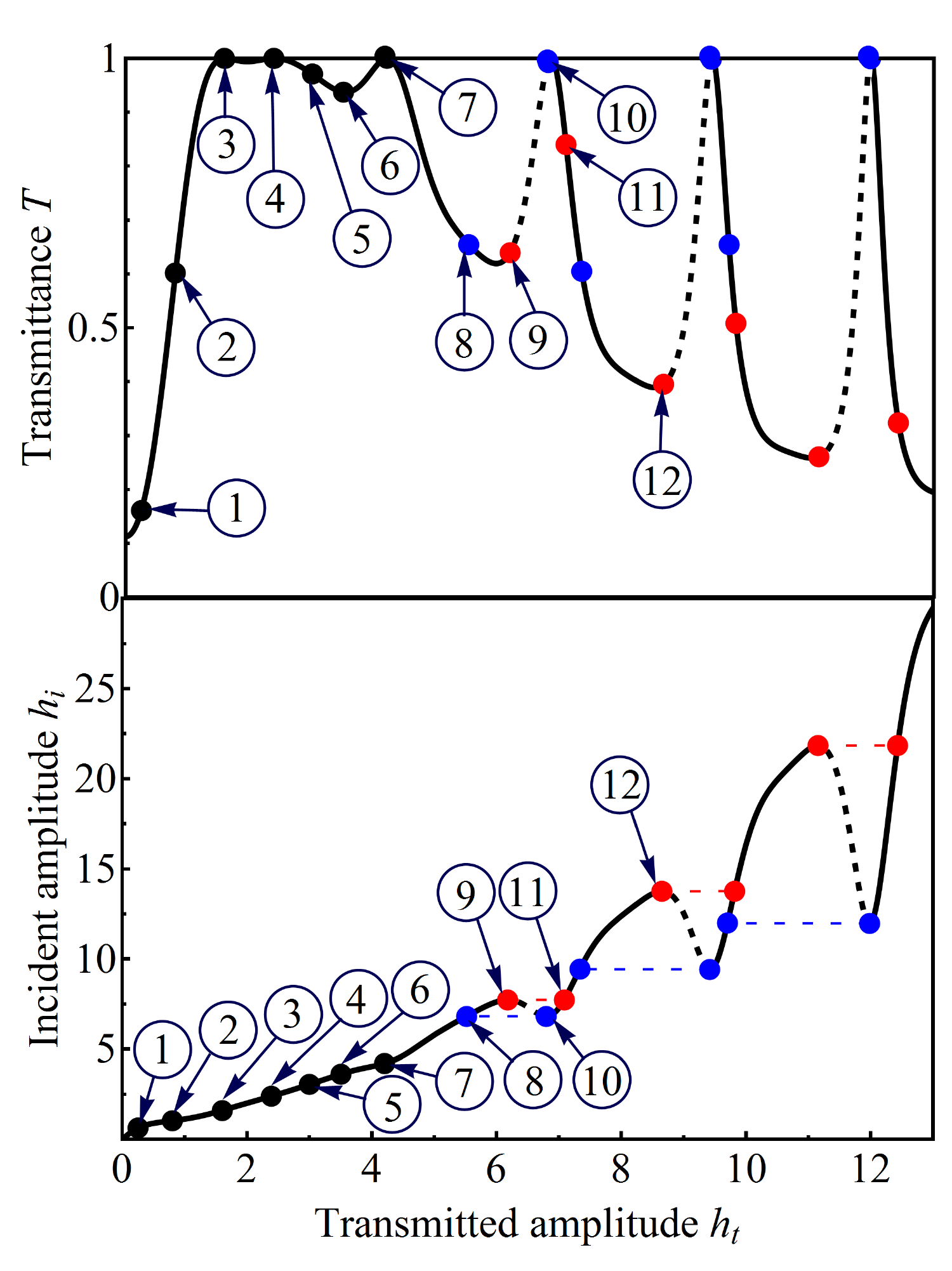}
\caption{(Color online) Solution of the inverse problem (the dependence of the incident wave amplitude $h_i$ and the transmittance $T=h_t^2/h_i^2$ on the transmitted wave amplitude $h_t$). The values of the parameters and the numbers of the points on the curves correspond to the numbers in Fig.~\ref{T(hi)_ab-1}. The curves show the nonmonotonic behavior of $h_i(h_t)$. This results in the multivalued dependence $T(h_i)$.} \label{fignonmon_ab}
\end{center}
\end{figure}

The dependence $U_{\rm eff} (a)$ illustrated in Fig.~\ref{mechanical_analogy_ab} shows that the potential energy is a single-valued function of $a$, although the dependence $T(h_i)$ is multivalued (see Fig.~\ref{T(hi)_ab-1}). This feature seems to be paradoxical, because the particle motion in any potential field is unambiguously defined by the initial conditions.  However, an assignment of the value of $h_i$ of the incident wave amplitude in Eqs.~(\ref{1ab}--\ref{3ab}) \emph{not always mean an imposition of definite initial conditions} for the particle motion. To demonstrate this, let us consider the inverse problem. We seek the amplitude $h_i$ of the incident wave that is necessary to obtain a definite amplitude $h_t$ of the transmitted wave. From Eqs.~\eqref{2ab} and \eqref{4ab}, we see that the value of $h_t$ defines unambiguously the angular momentum $L=h_t^2/\Gamma$ of the particle. According to the motion equations \eqref{from_sine-Gordon_ab-1}, \eqref{from_sine-Gordon_ab-2} and the boundary conditions Eqs.~(\ref{1ab}--\ref{3ab}), the dependence of $h_i$ on the amplitude $h_t$ of the transmitted wave and, correspondingly, the dependence of the transmittance
\[
T=h_t^2/h_i^2
\]
on $h_t$ are single-valued (see Fig.~\ref{fignonmon_ab}).  Nevertheless, \emph{the dependence $h_i(h_t)$ is nonmonotonic}. Therefore, the dependence $T(h_i)$ appears to be multiple-valued.

\section{Propagation of waves across superconducting layers}\label{c}

\begin{figure}[h]
\begin{center}
\includegraphics [width=14cm]{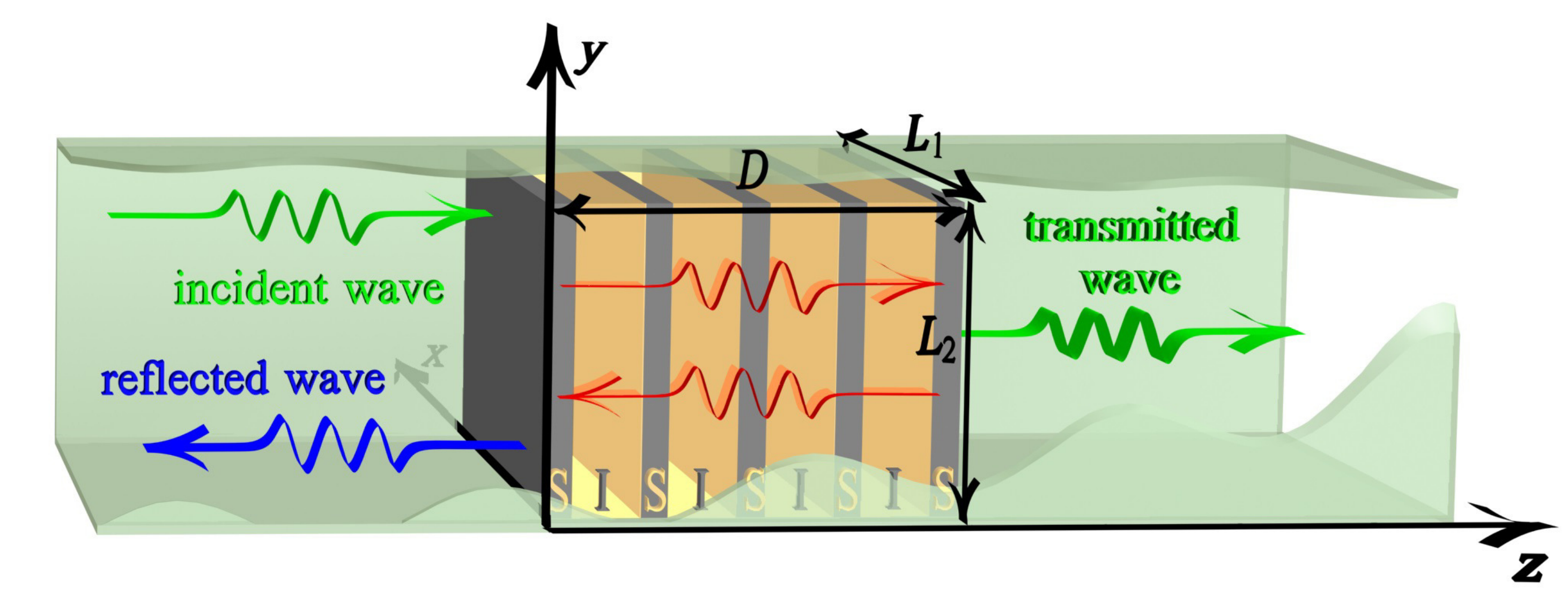}
\caption{(Color online) Schematic geometry for JPWs propagating in a  waveguide across the superconducting layers. Note that here S and I stand for superconducting and insulator layers, respectively. The green translucent layer (cut-off to show the sample inside) represents the walls of the waveguide.} \label{wavegC}
\end{center}
\end{figure}

Consider now the case when the electromagnetic wave propagates in the waveguide across the superconducting layers, i.e., the crystallographic \textbf{c}-axis of the superconductor and the $z$-axis are parallel to the waveguide axis (see Fig.~\ref{wavegC}). Let the incident extraordinary wave have the following polarization:
\begin{equation}\label{polarization1}
{\vec E} = \{E_x, E_y, E_z\}, \qquad {\vec H} = \{H_x, H_y, 0\}.
\end{equation}

In the first vacuum region (at $z<0$) the electromagnetic field can be represented as a sum of incident and reflected waves with amplitudes $H_i$ and $H_r$, respectively. There is only the transmitted wave of amplitude $H_t$ in the second vacuum region (at $z> D$). Using the Maxwell equations and boundary conditions (the equality of the tangential components of the electric field to zero on the waveguide walls), one can derive the electric and magnetic field components in the first and second vacuum regions:
\begin{gather}
E_x^{(v1)}=\frac{k_xk_z}{k^2}\left[H_i \sin(k_zz-\omega t)- H_r \sin(k_zz+\omega t+\alpha)\right]
\notag\\ \times\cos(k_xx)\sin(k_yy),
\notag\\
E_y^{(v1)}=\frac{k_yk_z}{k^2}\left[H_i \sin(k_zz-\omega t)- H_r \sin(k_zz+\omega t+\alpha)\right]
\notag\\ \times\sin(k_xx)\cos(k_yy),
\label{vacuum1}\\
H_x^{(v1)}=-\frac{k_y}{k}\left[H_i \sin(k_zz-\omega t)+ H_r \sin(k_zz+\omega t+\alpha)\right]
\notag\\ \times\sin(k_xx)\cos(k_yy),
\notag\\
H_y^{(v1)}=\frac{k_x}{k}\left[H_i \sin(k_zz-\omega t)+ H_r \sin(k_zz+\omega t+\alpha)\right]
\notag\\ \times\cos(k_xx)\sin(k_yy),
\notag
\end{gather}
\begin{gather}
E_x^{(v2)}=\frac{k_xk_z}{k^2}H_t \sin[k_z(z-D)-\omega t+\beta]\cos(k_xx)\sin(k_yy) ,
\notag\\
E_y^{(v2)}=\frac{k_yk_z}{k^2} H_t \sin[k_z(z-D)-\omega t+\beta]\sin(k_xx)\cos(k_yy),
\label{vacuum2}\\
H_x^{(v2)}=-\frac{k_y}{k}H_t \sin[k_z(z-D)-\omega t+\beta]\sin(k_xx)\cos(k_yy) ,
\notag\\
H_y^{(v2)}=\frac{k_x}{k}H_t \sin[k_z(z-D)-\omega t+\beta]\cos(k_xx)\sin(k_yy).
\notag
\end{gather}
Here
\begin{equation}
k_x=\frac{\pi n_1}{L_1},\quad k_y=\frac{\pi n_2}{L_2},\quad k_z= \big(k^2-k_x^2-k_y^2\big)^{1/2},
\end{equation}
$n_1$ and $n_2$ are nonnegative integer numbers that define propagating modes in the waveguide (they cannot be equal to zero simultaneously), $\alpha$ and $\beta$ are the phase shifts of the reflected and transmitted waves. We do not write the equation for the $z$ component $E_z$ of the electric field  since it is not used further in the text.

For the region occupied by the layered superconductor, we seek a solution of Eq.~(\ref{sine-Gordon}) in the form of a mode,
\[
\varphi (x, y, z, t) = a(z)|1-\Omega^2|^{1/2}
\]
\begin{equation}\label{varphi}
\times \sin(k_xx) \sin(k_yy) \sin[\eta(z)-\omega t],
\end{equation}
with the $z$-dependent amplitude $a(z)$ and phase $\eta (z)$.

We introduce the dimensionless coordinate $\zeta$ and the normalized length of the sample as
\begin{equation}\label{zeta-delta}
\zeta=\frac{\bar{\kappa} z}{\lambda_{ab}},\qquad \bar{\delta}=\frac{\bar{\kappa} D}{\lambda_{ab}},
\end{equation}
where
\begin{equation}
\bar{\kappa}=\lambda_c \frac{k_{||}}{|1-\Omega^2|^{1/2}},\quad k_{||}=\left(k_x^2+k_y^2\right)^{1/2}.
\end{equation}
Substituting the phase difference $\varphi$ in the form of Eq.~\eqref{varphi} into Eq.~\eqref{sine-Gordon}, we obtain two differential equations for the phase $\eta(\zeta)$ and the amplitude $a(\zeta)$,
\begin{equation}
\eta^\prime(\zeta)=-\frac{L}{h^2(\zeta)},
\label{from_sine-Gordon-1}
\end{equation}
\begin{equation}
h^{\prime\prime}(\zeta)=a(\zeta)+\frac{L^2}{h^3}+\frac{h(\zeta)}{\bar{\kappa}^2},
\label{from_sine-Gordon-2}
\end{equation}
where $L$ is an integration constant, prime denotes derivation over $\zeta$, and
\begin{equation}\label{h(a)}
h(\zeta)=-{\rm sign} (\Omega-1) a(\zeta) - \frac{9}{128} a^3(\zeta).
\end{equation}
Note that the cutoff frequency for linear waves propagating \emph{across} the superconducting layers coincides with the Josephson plasma frequency $\omega_J$. Therefore, for the geometry considered in this section, the transmittance of the superconducting slab is sensitive to the frequency detuning $(\Omega-1)$.

Equations~\eqref{s-fields}, \eqref{varphi}, and  \eqref{h(a)} allow one to express the electromagnetic field inside the slab via the functions $h(\zeta)$ and $\eta(\zeta)$. The tangential components of the electric and magnetic fields are:
\begin{gather}
E_x^s =-\mathcal{H}_0 \bar{\Gamma} \frac{k_xk_z}{k} \frac{|1-\Omega^2|}{k_{||}\bar{\kappa}}
\notag\\
\times  \left\{h(\zeta) \cos[\eta(\zeta)-\omega t]\right\}^\prime\cos(k_xx) \sin(k_yy),
\notag\\
E_y^s =-\mathcal{H}_0 \bar{\Gamma} \frac{k_yk_z}{k} \frac{|1-\Omega^2|}{k_{||}\bar{\kappa}}
\notag\\
\times \left\{h(\zeta)\cos[\eta(\zeta)-\omega t]\right\}^\prime\sin(k_xx) \cos(k_yy),
\label{tan_components}\\
H_x^s = \mathcal{H}_0 k_y \frac{|1-\Omega^2|}{k_{||}\bar{\kappa}} h(\zeta)
\notag\\
\times \sin[\eta(\zeta)-\omega t]\sin(k_xx) \cos(k_yy) ,
\notag\\
H_y^s =-\mathcal{H}_0 k_x \frac{|1-\Omega^2|}{k_{||}\bar{\kappa}} h(\zeta)
\notag\\
\times \sin[\eta(\zeta)-\omega t]\cos(k_xx) \sin(k_yy)
\notag
\end{gather}
with
\begin{equation}
\bar{\Gamma}=\frac{k^2 \lambda_{ab}}{k_z}\bar{\kappa}.
\end{equation}

Matching the fields given by Eqs.~\eqref{tan_components} for the superconductor and Eqs.~\eqref{vacuum1}, \eqref{vacuum2} for the vacuum regions, we  obtain the following three equations for the amplitudes $a(0)$, $a(\bar{\delta})$ and their derivatives on both edges of the layered superconductor, as well as the expression for the amplitude of the transmitted wave:
\begin{gather}
\bar{\Gamma}^2 [h^\prime(0)]^2+\left[h(0)+\frac{\bar{\Gamma} L}{h(0)} \right]^2=4 h_i^2,
\label{1}\\
h^2(\bar{\delta})=\bar{\Gamma} L,
\label{2}\\
h^\prime(\bar{\delta})=0,
\label{3}\\
h_t^2=h^2(\bar{\delta}).
\label{4}
\end{gather}
Here
\begin{equation}\label{normalization}
h_m=\frac{H_m}{\mathcal{H}_0} \frac{k_{||}\bar{\kappa}}{k |1-\Omega^2|},\qquad m=i,t
\end{equation}
are the normalized amplitudes of the incident ($m=i$) and transmitted ($m=t$) waves.
These equations, together with Eqs.~\eqref{from_sine-Gordon-1} and \eqref{from_sine-Gordon-2}, determine the integration constant $L$ as a function of the amplitude $h_i$ of the incident wave. Similarly to the previous section,  the constant $L$ defines directly the transmittance $T$ of the superconducting slab,
\begin{equation}\label{R}
T=\frac{h^2(\bar{\delta})}{h_i^2}=\frac{\bar{\Gamma} }{h_i^2}L.
\end{equation}

\begin{figure}
\begin{center}
\includegraphics*[width=14cm]{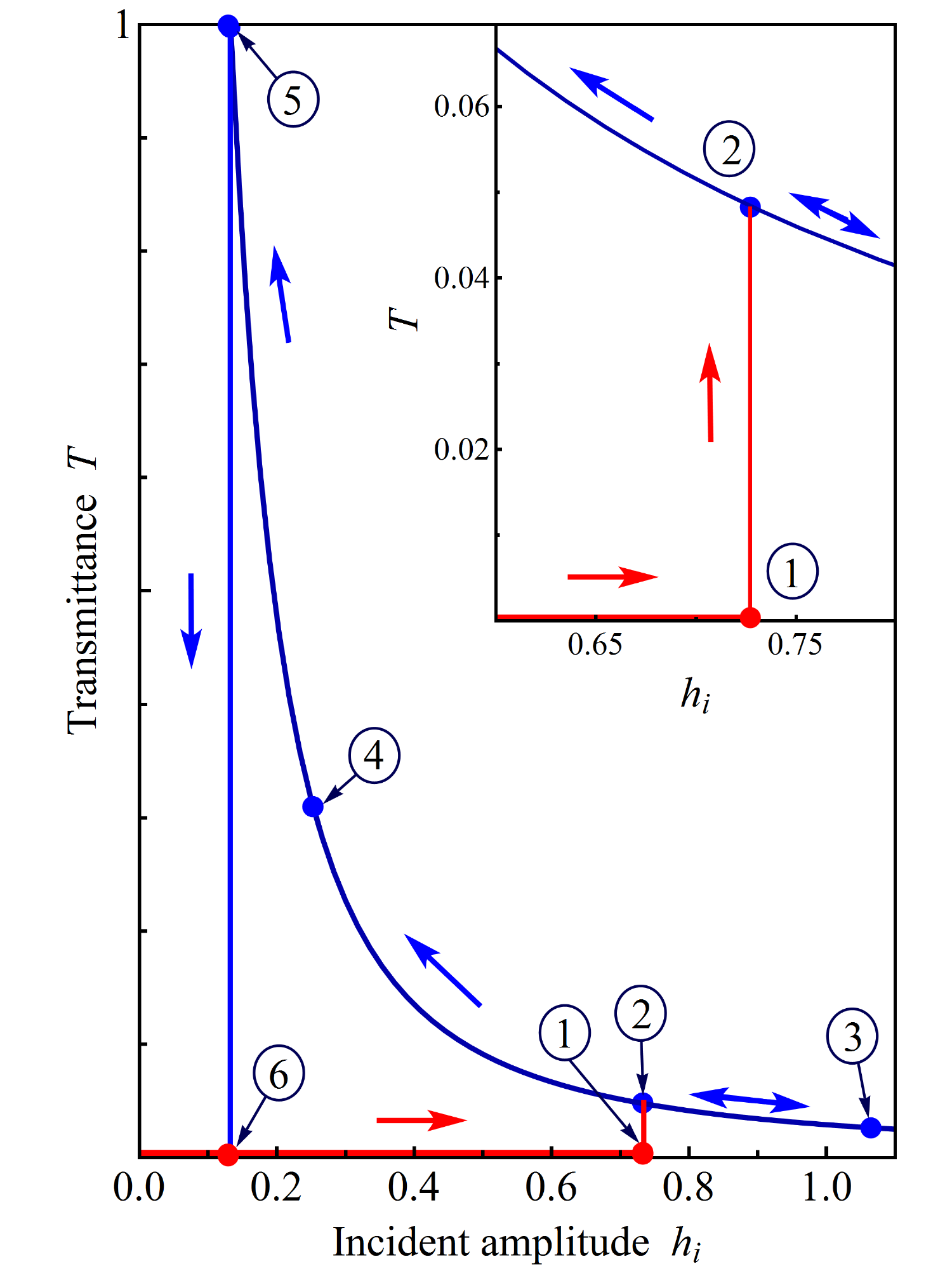}
\caption{(Color online) Dependence of the transmittance $T$ on the amplitude $h_i$ of the incident wave  for negative frequency detuning, $\Omega-1= - 5 \cdot 10^{-5}$. Arrows show the change of the transmittance when changing $h_i$. The inset shows the enlarged region near the points 1 and 2. The values of the parameters used here are: $\bar{\delta}=2$, $\lambda_c=4\cdot 10^{-3}$~cm, $\lambda_{ab}=2000$~\AA, $\omega_J/2\pi=0.3$~THz, $n_1=n_2=1$, and $L_1=L_2=0.1$~cm.} \label{T(hi)_omega<1}
\end{center}
\end{figure}

Figure~\ref{T(hi)_omega<1} shows the numerically-calculated hysteretic dependence of the transmittance $T$ on the amplitude $h_i$ of the incident wave for the case of negative detuning, $\Omega - 1 <0$.
The red curve close to the abscissa in Fig.~\ref{T(hi)_omega<1} shows the \emph{low-amplitude} branch of the $T(h_i)$ dependence. The low-amplitude waves cannot propagate across the superconducting layers at $\Omega < 1$ and, therefore, the transmittance is exponentially small on this branch.  When increasing the amplitude $h_i$ of the incident wave, the jump from the low-amplitude branch of the $T(h_i)$ dependence to the high-amplitude branch occurs at $h_i=(128/243)^{1/2}$. The high-amplitude branch is shown by the blue solid curve in Fig.~\ref{T(hi)_omega<1}. The high-amplitude solutions describe nonlinear Josephson plasma waves that can propagate in the layered superconductor for even  negative frequency detunings. Changing the amplitude $h_i$ one can control the relation between the wavelength and the length of the slab, and the transmittance varies from nearly zero to one depending on this relation. Thus, one can obtain total transparency of the slab choosing the optimal value of the amplitude $h_i$.

\begin{figure}
\begin{center}
\includegraphics*[width=14cm]{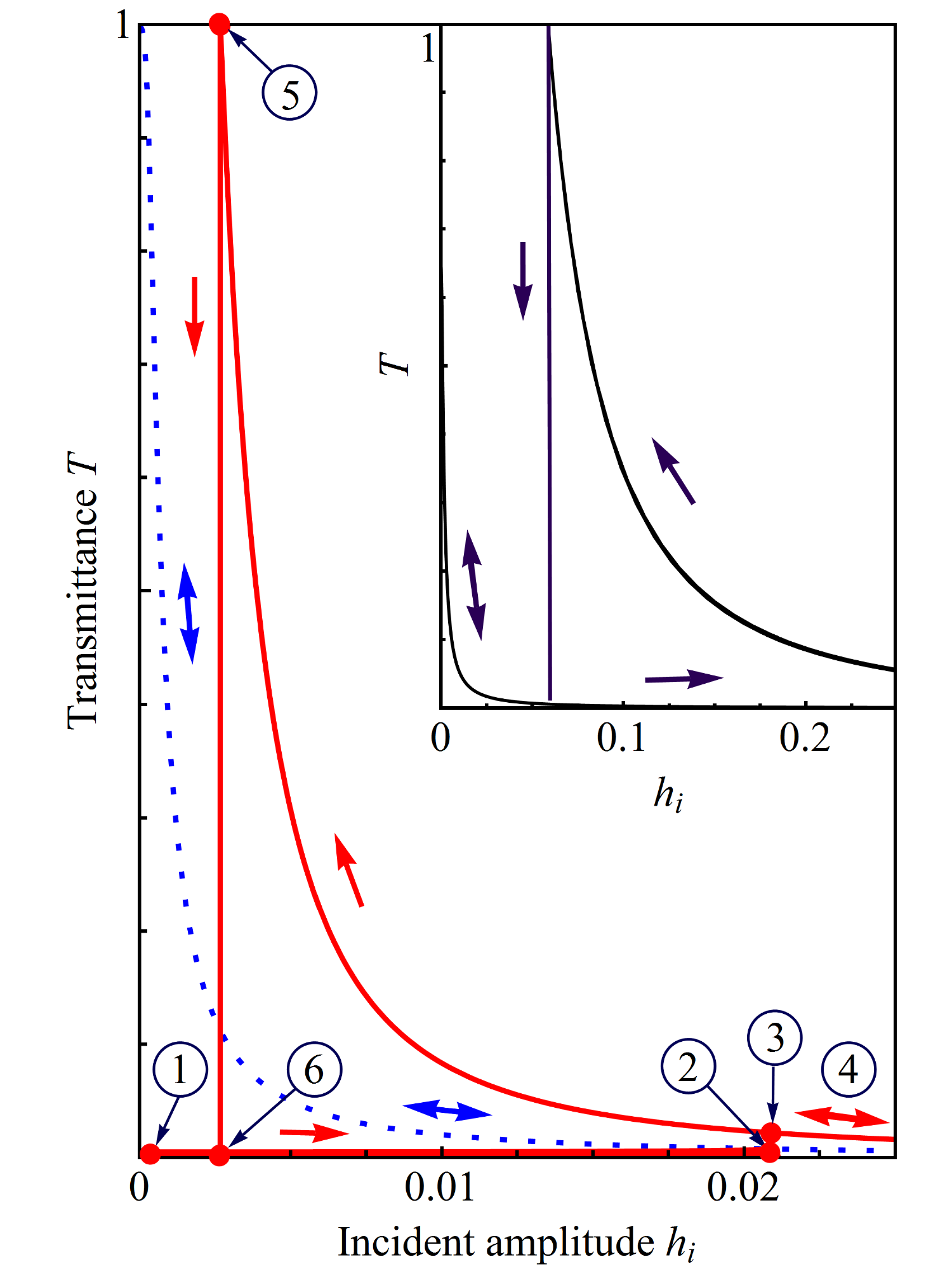}
\caption{(Color online) Dependence of the transmittance $T$ on the amplitude $h_i$ of the incident wave for different positive values of the frequency detuning: $\Omega-1= 4.97\cdot 10^{-3}$, or $\bar{\delta} /\pi = 1.21$ (dotted curve); $\Omega-1=4.74\cdot 10^{-3}$, or $\bar{\delta} /\pi = 1.24$ (solid curve); $\Omega-1=1.64\cdot 10^{-3}$, or $\bar{\delta} /\pi = 2.11$ (inset). Arrows show the change of the transmittance when changing the amplitude $h_i$ of the incident wave. The sample length is $D=4.3 \cdot 10^{-5}$~cm. Other parameters are the same as in Fig.~\ref{T(hi)_omega<1}.}
\label{T(hi)_omega>1}
\end{center}
\end{figure}

Figure~\ref{T(hi)_omega>1} shows the $T(h_i)$ dependence for the case of positive detunings, $\Omega - 1 >0$.
In this case, even linear JPWs can propagate in the superconductor across the layers. Therefore, the linear transmittance can vary from nearly zero to one depending on the relation between the length of the slab and the wavelength. The analysis of Eqs.~(\ref{from_sine-Gordon-1}--\ref{h(a)}), (\ref{1}--\ref{4}), and~\eqref{R} shows that the dependence $T(h_i)$ is single-valued if the frequency detuning exceeds some threshold value. For small $\bar{\Gamma}$, this threshold value is defined by the following asymptotic equation:
\begin{equation}\label{thr}
(\Omega_{\rm thr}-1)\, \approx \, \left(\frac{D\lambda_c k_{\|} }{\sqrt{2}\pi\lambda_{ab}}\right)^2.
\end{equation}
Such a reversible $T(h_i)$ dependence is shown by the dotted curve in Fig.~\ref{T(hi)_omega>1}. The hysteresis in the $T(h_i)$ dependence exists for frequencies smaller than the threshold value, for $\Omega < \Omega_{\rm thr}$. In this case, the complete transparency of the sample can be observed if the incident wave amplitude $h_i$ is first increased and then decreased.

Note that plots in Figs.~\ref{T(hi)_omega<1} and \ref{T(hi)_omega>1} are very similar to $T(h_i)$ curves obtained in Ref.~\onlinecite{nl4}  for the case of an infinite slab. It is not surprising because the statements of these two problems differ in not very important details. Namely, the nonlinear waves \emph{running } in the infinite slab at some angle  with respect to the superconducting layers were considered in Ref.~\onlinecite{nl4}, whereas  the nonlinear waves \emph{standing}  in the $x$- and $y$-directions and propagating only across the layers are studied in this section. Therefore, the sets of equations for finding the transmittance in these two problems differ in the numerical coefficients only.

\section{Conclusion}

In this paper, we have studied theoretically the phenomenon of self-induced transparency of finite-length slabs of layered superconductors placed in a waveguide with ideal walls. This phenomenon can be observed for two geometries: when the superconducting layers are parallel to the waveguide axis or perpendicular to them. We show that the transmittance of a superconducting slab is very sensitive to the amplitude of the incident wave due to the nonlinear dependence of the Josephson \textbf{c}-axis current on the interlayer phase difference of the order parameter. A very interesting feature of the predicted phenomenon is the hysteretic behavior of the $T(h_i)$  dependence.  It is important to note that the tunable transmittance can vary over a wide range, from nearly zero to one, when changing the incident wave amplitude, even in the case of weak nonlinearity when the interlayer phase difference is small.

\section{Acknowledgements}

We gratefully acknowledge partial support from the ARO, JSPS-RFBR Contract No.~12-02-92100, Grant-in-Aid for Scientific Research (S), MEXT Kakenhi on Quantum Cybernetics, the JSPS-FIRST program, Ukrainian State Program on Nanotechnology, and the Program FPNNN of the NAS of Ukraine (grant No~9/11-H).


\begin{thebibliography}{99}

\bibitem{Kl-Mu} R.~Kleiner, F.~Steinmeyer, G.~Kunkel, and P.~M\"{u}ller, \prl \textbf{68}, 2394 (1992).

\bibitem{Kl-Mu2} R.~Kleiner and P.~M\"{u}ller, \prb \textbf{49}, 1327 (1994).

\bibitem{Thz-rev} S.~Savel'ev, V.A.~Yampol'skii, A.L.~Rakhmanov, and F.~Nori, Rep. Prog. Phys. \textbf{73}, 026501 (2010).

\bibitem{rev2}X. Hu and S.-Z. Lin, Supercond. Sci. Technol. \textbf{23}, 053001 (2010).

\bibitem{surf} S.~Savel'ev, V.~Yampol'skii, and F.~Nori, \prl {\bf 95}, 187002 (2005);

\bibitem{neg-ref}  V.A.~Golick, D.V.~Kadygrob, V.A.~Yampol’skii, A.L.~Rakhmanov, B.A.~Ivanov, and
F.~Nori, Phys. Rev. Lett. \textbf{104}, 187003 (2010).

\bibitem{PhysRevB1} V.A.~Yampol'skii, A.V.~Kats, M.L.~Nesterov, A.Yu.~Nikitin,
T.M.~Slipchenko, S.~Savel'ev, and F.~Nori, \prb {\bf 76}, 224504 (2007).

\bibitem{agr} V.M.~Agranovich and D.L.~Mills, \textit{Surface Polaritons}, edited by V.M.~Agranovich and D.L. Mills (North-Holland, Amsterdam, 1982).

\bibitem{wood} H.~Raether, {\it Surface Plasmons} (Springer-Verlag, New York, 1988).

\bibitem{Petit} R.~Petit, {\it Electromagnetic Theory of Gratings} (Springer, Berlin, 1980).

\bibitem{negref} A.L.~Rakhmanov,  V.A.~Yampol'skii, J.A.~Fan, F.~Capasso, and F.~Nori, Phys. Rev. B \textbf{81}, 075101 (2010).

\bibitem{filter} V.A.~Yampol’skii, S.~Savel’ev, O.V.~Usatenko, S.S.~Mel’nik, F.V.~Kusmartsev, A.A.~Krokhin, and F.~Nori, Phys. Rev. B {\bf 75}, 014527 (2007).

\bibitem{nl1} S.~Savel’ev, A.L.~Rakhmanov, V.A.~Yampol’skii, and F.~Nori, Nature Phys. {\bf 2}, 521 (2006).

\bibitem{nl2} V.A.~Yampol’skii, S.~Savel’ev, A.L.~Rakhmanov, and F.~Nori, Phys. Rev. B {\bf 78}, 024511 (2008).

\bibitem{nl3} S.~Savel’ev, V.A.~Yampol’skii, A.L.~Rakhmanov, and F.~Nori, Phys. Rev. B {\bf 75}, 184503 (2007).

\bibitem{nl4} S.S.~Apostolov, Z.A.~Maizelis, M.A.~Sorokina, V.A.~Yampol’skii, and F.~Nori, Phys. Rev. B {\bf 82}, 144521 (2010).

\bibitem{sine-gord} S.~Sakai, P.~Bodin, and N.F.~Pedersen, J. Appl. Phys. {\bf 73}, 2411 (1993).

\bibitem{SG2} S.N.~Artemenko and S.V.~Remizov, JETP Lett. {\bf 66}, 811 (1997).

\bibitem{SG3} S.N.~Artemenko and S.V.~Remizov, Physica C {\bf 362}, 200 (2001).

\bibitem{SG4} Ch.~Helm, J.~Keller, Ch.~Peris, and A.~Sergeev, Physica C {\bf 362}, 43 (2001).

\bibitem{SG5} Ju.H.~Kim and J.~Pokharel, Physica C {\bf 384}, 425 (2003).

\bibitem{helm3} Ch.~Helm and L.N.~Bulaevskii, \prb {\bf 66}, 094514 (2002).

\end{thebibliography}
\end{document}